\documentclass[a4paper,12pt]{article}

\usepackage{graphicx}
\usepackage{epsfig}
\usepackage[american]{babel}

\begin{document}
\title{Scaling at the onset of chaos in a network of logistic maps
with two types of global coupling}

\author{A.S.Ivanova, S.P.Kuznetsov}

\maketitle\begin{center} \emph{Institute of Radio-Engineering and
Electronics of RAS, \\ Zelenaya 38, Saratov, 410019,
Russia}\end{center}

\maketitle\begin{center} \emph{Saratov State University, \\
Astrakhanskaya 83, Saratov, 410026, Russia}\end{center}

\maketitle
\begin{abstract}
We study a network of logistic maps with two types of global
coupling, inertial and dissipative. Features of the clusterization
process are revealed and compared, which are associated with
presence of each type of couplings. For the parameter region near
the onset of chaos we outline scaling properties in dynamical
behavior of the network and illustrate them on the Kaneko phase
diagrams.
\end{abstract}

\newpage

Recently a notable interest is attracted to systems of globally
coupled elements [1-4]. In particular, such models are discussed
in context of description of neural networks, multi-modal lasers,
electronic and chemical systems. They are regarded as perspective
for nontrivial application in information processing to reproduce,
at least in some respects, analogous processes in biological
systems.

One particular class of networks, rather simple but possessing
many nontrivial peculiarities in the dynamical behavior, was
introduced by Kaneko [1]. It is a set of logistic maps, each
coupled to each other with equal strength. Studies of Kaneko and
other authors revealed a reach phenomenology of dynamics of this
model. In particular, a phenomenon called clusterization has been
discovered -- this is a spontaneous formation of groups of
elements in process of dynamics of the network with instantaneous
states exactly coinciding for elements relating to each group
(cluster). A possibility of coexistence of attractors with
different number of clusters at the same parameter values of the
network gives an opportunity for information storage; control of
them may be used for information processing.

In the original work of Kaneko and in many subsequent researches,
a special concrete type of coupling was assumed: it tends to
equalize instantaneous states of the interacting elements and is
called the dissipative coupling [5,6]. However, it is not the only
possible type of interaction of the elements. As found from the
renormalization group (RG) analysis of the onset of chaos in
coupled period-doubling maps and in coupled map lattices, in
general case a weak coupling introduced by smooth functions of
state variables is a combination of \textit{two} relevant types of
coupling -- inertial and dissipative [5-8]. Both types of coupling
possess distinct properties in respect to the RG transformation
and manifest different scaling regularities at the onset of chaos
in coupled logistic maps. A necessity of consideration of globally
coupled models with two types of coupling was mentioned e.g. in
Refs.[ 9, 10], but neither convenient models, nor detailed studies
of features of dynamics were reported there.

In the present paper we introduce a model of a network of logistic
maps with dissipative and inertial global coupling. Some features
of dynamics and structure of phase diagrams will be discussed and
compared for the networks with these types of coupling. In
particular, we outline properties of universality and scaling in
dynamics of the globally coupled maps near the onset of chaos. Due
to richness of dynamical phenomena, this region may be especially
interesting for applications in the information processing.

Let us consider a network of coupled logistic maps governed by the
following equations:

\begin{equation} \label{eq1} x_{n + 1} (i) = (1 - \varepsilon _{2}
)f(x_{n} (i)) - \varepsilon _{1} x_{n} (i) + {\frac{{\varepsilon
_{1}} }{{N}}}{\sum\limits_{j = 1}^{N} {x_{n} (j)} } +
{\frac{{\varepsilon _{2}} }{{N}}}{\sum\limits_{j = 1}^{N} {f(x_{n}
(j))} }{\rm ,} \end{equation}

\noindent where $f(x) = 1 - \lambda x^{2}$ is a nonlinear function
corresponding to the logistic map, $n$ designates the discrete
time, index $i$ enumerates elements of the network, $N$ is the
total number of the elements,\textit{ $\varepsilon $} and
\textit{$\varepsilon $} are parameters of coupling.

Note that two last terms in Eq. (\ref{eq1}) do not depend on the
index $i$, in other words, they are the same for all elements of
the network. Hence, they may be regarded as two mean fields
responsible for two types of global coupling:

 $F_{n}^{(\ref{eq1})} = N^{ - 1}{\sum\limits_{j = 1}^{N} {x_{n}} } (j)$ and $F_{n}^{(2)}
= N^{ - 1}{\sum\limits_{j = 1}^{N} {f(x_{n}} } (j))$. (2)

As follows from previous works on coupled logistic maps [5-10],
the term with $f$($x$) in Eq.(\ref{eq1}) and the mean field
$F^{(2)}$ represent the pure dissipative coupling. The linear term
and the term $F^{(\ref{eq1})}$ correspond to a combination of
inertial and dissipative coupling, but the dissipative
contribution is rather small. Quantitatively, the coefficients
$\varepsilon _{1} $ and $\varepsilon _{2} $ are expressed via the
coefficients of inertial and dissipative coupling $\varepsilon
_{I} $ and $\varepsilon _{D} $ as follows [6,8]:

\begin{equation} \label{eq2} \varepsilon _{I} = \varepsilon _{1}
,\,\,\;\varepsilon _{D} = \varepsilon _{2} - 0.088\varepsilon _{1}
{\rm .} \end{equation}

In the system with both types of coupling we expect to observe the
same phenomenon of clusterization discovered by Kaneko for the
dissipative case. Indeed, the main reason --coincidence of
conditions of motion for elements with equal instantaneous states
is valid irrespectively to the number of mean fields.

To analyze dynamics of the globally coupled network in dependence
on the parameters let us follow the approach of Kaneko based on
the concept of \textit{phases}.

Been given a point in the parameter space (\textit{$\lambda  \quad
\varepsilon $, $\varepsilon $}) we consider an ensemble of
identical independent networks, each with its own, randomly chosen
initial conditions. Performing a sufficiently large number of
iterations, we estimate a number of clusters been formed in each
representative of the ensemble. Accounting statistics of the final
states, we classify the phase as coherent (one-cluster state
dominates in the ensemble), ordered (states with few clusters are
of the highest probability), partially ordered (clusters with
small and large numbers of elements appear with comparable
probabilities), turbulent (number of clusters is of order of the
total number of elements in the network). On the phase diagrams of
Fig.1 the respective phases are indicated by symbols C, O, PO, and
T. Panel (a) relates to the case of inertial coupling, this is
cross-section of the parameter space by plane (\textit{$\lambda $}
$\varepsilon _{1} = \varepsilon _{I} , \varepsilon _{2} =
0.088\varepsilon _{I} $). Panel (b) corresponds to dissipative
coupling and reproduce the diagram computed by Kaneko. It
represents a cross-section of the parameter space by plane
(\textit{$\lambda $} $\varepsilon _{1} = 0_{I} , \varepsilon _{2}
= \varepsilon _{D} $) and is presented here for comparison. (Here
we restrict ourselves by frames of a rough approach to the
definition of phases, as in the original work of Kaneko, because
it is sufficient to reach our aims and to demonstrate scaling.
However, it has been shown recently that detection of presence or
absence of clusterization, in a sense of exact coincidence of
instantaneous states of elements, especially in chaotic domain,
require much more subtle numerical technique and analysis [4].)

Observe that in the domain of subcritical dynamics of individual
elements the network with inertial coupling demonstrates more
reach dynamics than in the case of dissipative coupling. Presence
of phases O and PO indicates that formation of multiple clusters
and multistability takes place here, in domain of regular
dynamics. Perhaps, this implies that the networks with inertial
coupling should be considered as more perspective candidates for
applications in information processing than those with dissipative
coupling.

In Refs.[5-10] it was stated that coupled maps with inertial and
dissipative coupling near the onset of chaos demonstrate a
property of scaling. The network with global coupling may be
thought as a set of elements coupled pairwise. Hence, analogous
scaling regularities must be valid for the global coupling, and
the statement may be formulated as follows.

Let us suppose that at the parameters $\lambda  \quad \varepsilon
$, $\varepsilon $ we detect some phase for the ensemble of systems
(\ref{eq1}) with initial conditions randomly chosen from interval
${\left| {x} \right|} < C$. Then, for ensemble with random initial
conditions from interval ${\left| {x} \right|} < {{C}
\mathord{\left/ {\vphantom {{C} {{\left| {\alpha} \right|}}}}
\right. \kern-\nulldelimiterspace} {{\left| {\alpha} \right|}}}$
at the point $\lambda _{ñ} + {{(\lambda - \lambda _{ñ} )}
\mathord{\left/ {\vphantom {{(\lambda - \lambda _{ñ} )} {\delta}
}} \right. \kern-\nulldelimiterspace} {\delta} }$, ${{\varepsilon
_{I}} \mathord{\left/ {\vphantom {{\varepsilon _{I}}  {\alpha} }}
\right. \kern-\nulldelimiterspace} {\alpha} }$, ${{\varepsilon
_{D}} \mathord{\left/ {\vphantom {{\varepsilon _{D}}  {2}}}
\right. \kern-\nulldelimiterspace} {2}}$, the same kind of phase
will be observed, but with doubled time scale of motion. Here
$\lambda $ is the parameter value of the period-doubling
accumulation in an individual element of the network (the logistic
map), $\alpha $--2.5029\ldots and $\delta $4.6692\ldots are the
Feigenbaum universal constants.

Straightly speaking, the above statement is true only
asymptotically: smaller the neighborhood of the critical point
($\lambda $, 0, 0), more accurate the scaling property holds. But
actually it works very well, even in sufficiently large scales. It
may be seen from comparison of the phase diagrams of Fig 1 (a) and
(c), (b) and (d). The bottom pictures present results obtained
with rescaling of the parameters and initial conditions in
accordance with the formulated rules.

In, conclusion, we have introduced a model network of logistic
maps with two types of global coupling. We argued that the Kaneko
classification of phases suggested for the network of
dissipatively coupled maps remains valid for our generalized
model. In contrast to the dissipative coupling, the inertial one
ensures reach dynamics (multi-stability, states of multiple
clusters -- the phases O and PO) not only in supercritical
(chaotic) domain of the individual elements, but also in
subcritical domain and at the border of chaos, where the dynamics
is more regular. This circumstance indicates that the networks
with inertial coupling are more interesting for possible
applications in information processing than those with dissipative
coupling.

For the first time we have demonstrated scaling regularities in
phase diagrams of the models with global coupling. Note
significance of this observation. As follows from RG argumentation
developed earlier for coupled maps and for coupled map lattices,
the scaling property appears as an attribute of the universality
class. It means that analogous dynamical regimes and the same
regularities will be intrinsic not only to our logistic map model,
but for globally coupled networks composed of any other
period-doubling elements that relate to the Feigenbaum
universality class (say, of forced nonlinear dissipative
oscillators) [11,12]. So, the results are expected to be common
for globally coupled networks of different physical nature
(electronics, biology, economics, etc.).

\textit{The work was supported, in part, by RFBR (grant No
00-02-17509) and by Ministry for Industry, Science and Technology
of Russian Federation (State Contract No 40.020.1.1.1168).}

\vspace{10mm}
\noindent
FIGURE CAPTION
\vspace{5mm}

\noindent
Fig.1. Kaneko's phase diagrams in cross-section of
the parameter space of the globally coupled network of logistic
maps by planes of pure inertial coupling (a,c) and pure
dissipative coupling (b,d). The bottom pictures are built with the
use of scaling rules formulated in the main text. Observe
similarity of the diagrams (a) and (c), (b) and (d). The symbols
designate Kaneko's phases: C -- coherent, O -- ordered, PO --
partially ordered, T -- turbulent.


\begin{thebibliography}{99}

\bibitem{b1} Kaneko K. \textit{Clustering, Coding, Switching,
Hierarchical Ordering, and Control in Network of Chaotic
Elements}. Physica \textbf{D41}, 1990, No 2. p. 137.

\bibitem{b2} Kuznetsov A.P., Kuznetsov S.P. \textit{Critical dynamics of
coupled maps near the onset of chaos (A review).} Izvestija VUZov
-- Radiofizika, \textbf{34}, 1991, Nos 10-12, p.1079 (in Russian).

\bibitem{b3} Taborov A. V., Maistrenko Yu. L., Mosekilde E.
\textit{Partial Synchronization in a System of Coupled Logistic
Map.} International Journal of Bifurcation and Chaos,
\textbf{10}, 2000, No. 5, p. 1051.

\bibitem{b4} Pikovsky A., Popovich O., Maistrenko Yu. \textit{Resolving
clusters in chaotic ensembles of globally coupled identical
oscillators.} Phys.Rev.Lett., \textbf{87}, 2001, No 4, 044102.

\bibitem{b5} Kuznetsov S.P. \textit{Universality and similarity in
behavior of coupled Feigenbaum systems.} Izvestija VUZov --
Radiofizika, \textbf{28}, 1985, No 8, 991. (English translation:
Radiophysics and Quantum Electronics, \textbf{28}, 1985, 681.)

\bibitem{b6} Kuznetsov S. P. \textit{Renormalization group,
universality and scaling in dynamics of coupled map lattices.} In
book: \textit{Theory and applications of coupled map lattices}.
Edited by K. Kaneko: John Wiley \& Sons Ltd. 1993, p.50-93.

\bibitem{b7} Kook H., Ling F.H., Schmidt G. \textit{Universal behavior of
coupled nonlinear systems.} Phys.Rev. \textbf{A43}, No 6, 1991,
p.2700.

\bibitem{b8} Kuznetsov S.P. \textit{Universality and Scaling in
Two-Dimensional Coupled Map Lattices.} Chaos, Solitons and
Fractals, \textbf{2}, 1992, No 3, p.281.

\bibitem{b9} Kim S.-Y., Kook H. \textit{Period doubling in coupled maps.}
Phys.Rev. \textbf{E48}, No 2, 1993, p.785.

\bibitem{b10} Kim S.-Y., Kook H. \textit{Renormalization analysis of two
coupled maps.} Phys.Lett. \textbf{A178}, 1993, p.258.

\bibitem{b11} Feigenbaum M. J. \textit{Quantitative universality for a
class of nonlinear transformations.} J. Stat. Phys., \textbf{19},
1978, No 1, 25-52.

\bibitem{b12} Feigenbaum M. J. \textit{The universal metric properties of
nonlinear transformations.} J. Stat. Phys. \textbf{21}, 1979, No 6,
p.669.

\end{thebibliography}
\end{document}